\documentclass[amsmath, amssymb, amsfonts, twocolumn, footinbib]{revtex4-1}

\usepackage{amsmath}
\usepackage{graphicx}% Include figure files
\usepackage{dcolumn}% Align table columns on decimal point
\newcolumntype{d}[1]{D{.}{.}{#1}}

\usepackage{bm}% bold math
\usepackage{epstopdf}
\usepackage{graphicx}
\usepackage{stmaryrd}
\usepackage{tikz}
\usepackage{pgf}
\definecolor{lightblue}{RGB}{0,170,255}

\newcommand{\bt}[1]{{\textbf{#1}}}

\newcommand{\appropto}{\mathrel{\vcenter{
			\offinterlineskip\halign{\hfil$##$\cr
				\propto\cr\noalign{\kern2pt}\sim\cr\noalign{\kern-2pt}}}}}

\usepackage[normalem]{ulem}
\usepackage{color}

\begin{document}
	\title{Electrical control of the Zeeman spin splitting in two-dimensional hole systems}
	
	\author{E. Marcellina}
	\affiliation{School of Physics, The University of New South Wales, Sydney, Australia}
			
	\author{A. Srinivasan}
	\affiliation{School of Physics, The University of New South Wales, Sydney, Australia}
	
	\author{D. S. Miserev}
	\affiliation{School of Physics, The University of New South Wales, Sydney, Australia}
		
	\author{A. F. Croxall}
	\affiliation{Cavendish Laboratory, University of Cambridge, J. J. Thomson Avenue, Cambridge CB3 0HE, United Kingdom}
	
	\author{D. A. Ritchie}
	\affiliation{Cavendish Laboratory, University of Cambridge, J. J. Thomson Avenue, Cambridge CB3 0HE, United Kingdom}
	
	\author{I. Farrer}
	\affiliation{Cavendish Laboratory, University of Cambridge, J. J. Thomson Avenue, Cambridge CB3 0HE, United Kingdom}

	\author{O. P. Sushkov}
	\affiliation{School of Physics, The University of New South Wales, Sydney, Australia}

	\author{Dimitrie Culcer}
	\affiliation{School of Physics, The University of New South Wales, Sydney, Australia}
	
	\author{A. R. Hamilton}
	\affiliation{School of Physics, The University of New South Wales, Sydney, Australia}	
\date{\today}

\begin{abstract}
Semiconductor holes with strong spin-orbit coupling allow all-electrical spin control, with broad applications ranging from spintronics to quantum computation. Using a two-dimensional hole system in a GaAs quantum well, we demonstrate a new mechanism of electrically controlling  the Zeeman splitting, which is achieved through altering the hole wave vector $k$. We find a threefold enhancement of the in-plane $g-$factor $g_{\parallel}(k)$. We introduce a new method for quantifying the Zeeman splitting from magnetoresistance measurements, since the conventional tilted field approach fails for two-dimensional systems with strong spin-orbit coupling. Finally, we show that the Rashba spin-orbit interaction suppresses the in-plane Zeeman interaction at low magnetic fields. The ability to control the Zeeman splitting with electric fields opens up new possibilities for future quantum spin-based devices, manipulating non-Abelian geometric phases, and realising Majorana systems in $p-$type superconductor systems.
\end{abstract}	
	
\maketitle
	
The spin-orbit interaction couples a particle's spin to its motion, as described by the Hamiltonian $H_{\mathrm{SO}} = \frac{1}{2} \mu_B \boldsymbol{\sigma.B}_{\mathrm{eff}}(\boldsymbol{k})$, where $\boldsymbol{\sigma}$, $\mu_B$, $\boldsymbol{B}_{\mathrm{eff}}$, and $\boldsymbol{k}$  represent the Pauli matrices, the Bohr magneton, the effective spin-orbit magnetic field, and the motion wave vector, respectively \cite{Winkler2003}. The effective magnetic field $\boldsymbol{B}_{\mathrm{eff}}$ emerges from a relativistic transformation that occurs when a particle with spin $\boldsymbol{\sigma}$ is moving with wave vector $\boldsymbol{k}$ with respect to an electric field $\boldsymbol{F}$. The spin-orbit interaction has applications in spintronic devices and spin-based quantum computing, which rely on controlling spin via external electric fields $\boldsymbol{F}$ \cite{Datta1990,Zutic2004,Loss1998,Wolf2001,Awschalom2013,Nowack2007,Bulaev2007,NadjPerge2010,Pribiag2013,Maurand2016,Salfi2016,Hung2017}.

Here we report a new mechanism for electrically manipulating spin through the Zeeman interaction in two-dimensional (2D) holes. Unlike in electrons, the Zeeman splitting in holes is highly anisotropic, with the out-of-plane $g-$factor $g_{zz}$ much larger than the in-plane one ($g_{zz} \gg g_{\parallel}$), and $g_{\parallel}$ strongly dependent on $k$. The ability to electrically control  $g_{\parallel}$ is not only valuable for applications in spintronics and quantum computing, but also for engineering non-Abelian geometric phases  \cite{Budich2012,Li2016}. Moreover, in a hybrid semiconductor-superconductor system that can host Majorana fermions \cite{Alicea2011,Lutchyn2011,Mao2012,Mourik2012,Rokhinson2012,Vaitiekenas2017,Liang2017}, a high tunability of the $g-$factor is desirable as it is then possible to use a magnetic field to drive the system from the trivial to the topological regime without quenching the superconductivity needed to support the Majorana mode.
	
\begin{figure*}
	\centering
		\includegraphics[scale=1.07]{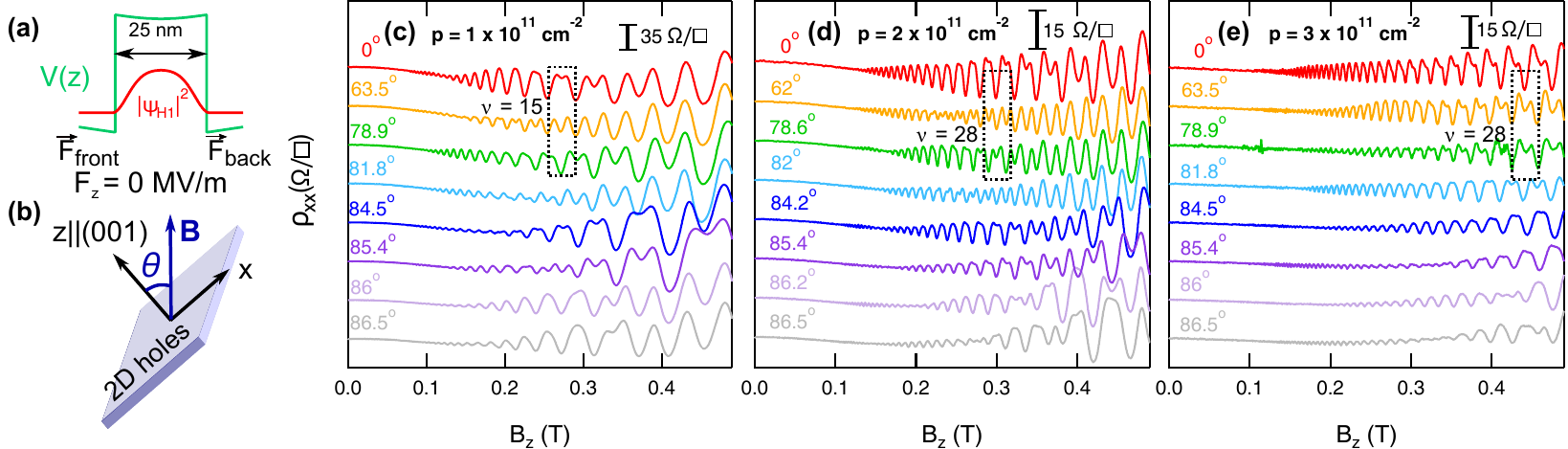}
		\caption{(a) Schematic of the GaAs hole quantum well used in this work, where $\psi_{H1}$ is the heavy hole wave function and $V(z)$ is the confinement potential controlled via electric fields $\protect\overrightarrow{F}_{\mathrm{front}}$ and $\protect\overrightarrow{F}_{\mathrm{back}}$ applied using the front and back gates, respectively. Here, the total electric field $\protect\overrightarrow{F} = F_z \hat{z} \equiv (\protect\overrightarrow{F}_{\mathrm{front}} + \protect\overrightarrow{F}_{\mathrm{back}})/2 = 0$. (b) The device is tilted at an angle $\theta$ with respect to the applied magnetic field $\bt{B}$. (c)-(e) Magnetoresistance oscillations $\rho_{xx}$ of the 2D holes in a symmetric quantum well at various tilt angles for three hole densities $p$. Due to the in-plane magnetic field, a resistance minimum gradually evolves into a maximum or vice versa, for example at $p = 1 \times 10^{11}$ cm$^{-2}$ and $\nu = 15$ for angles $\theta = 0^{\circ},~63.5^{\circ},~78.9^{\circ}$ (see dashed box). (d)-(e) The effect of the in-plane magnetic field on the resistance becomes more pronounced as the 2D hole density increases: the transition between a resistance minimum and maximum occurs at smaller angles at higher densities (see dashed box). The traces have been offset for clarity.}
	\label{fig:tilted_field_shubs}
\end{figure*}

Previous methods of tuning the $g-$factor in 2D systems relied on shifting the wave function from one material to another, either by pushing it across a heterointerface, or by using a graded composition quantum well \cite{Salis2001}. Here we adopt a different approach, in which the $g-$factor is controlled not by shifting the wave function but by tuning the Fermi wave vector $k_F$ of a 2D hole system. Owing to the spin$-3/2$ nature of the 2D holes \cite{Luttinger1956}, increasing $k$ enhances the mixing between the occupied heavy hole and unoccupied light hole subbands, which then dramatically alters the in-plane $g-$factor $g_{\parallel}$. It is difficult to detect this variation of $g_{\parallel}$ with optical methods, since these detect bound excitons with a small and fixed $k$ \cite{Syperek2007,Kugler2009,Gradl2014}. Instead we introduce a new approach based on magnetotransport in crossed magnetic fields, which shows that the spin splitting is linear in applied in-plane magnetic field, and can be varied by $300\%$ as $k_F$ is increased.

Our experiment was performed using an undoped 25 nm GaAs quantum well sandwiched between 300 nm Al$_{0.33}$Ga$_{0.67}$As layers and grown along the (001) direction. Metal top and back gates supplied the 2D carriers and allowed the quantum well symmetry to be tuned arbitrarily \cite{Croxall2013}. The magnetotransport measurements were performed in a dilution refrigerator at a base temperature of $\sim$ 30 mK. The sample was mounted on an in-situ rotation system with an accuracy of $\pm0.01^{\circ}$ \cite{Yeoh2010}. Electrical measurements were carried out in the standard 4-terminal configuration with a source-drain current of 5 nA and a lock-in frequency of 17 Hz. The maximum 2D hole mobility was $1.5 \times 10^{6}$ cm${^2}$ V$^{-1}$ s$^{-1}$ at density $p = 2.6 \times 10^{11}$ cm$^{-2}$. 

%\cite{Dresselhaus1955,Coleridge1989,Winkler2000,Habib2004,Keppeler2002,Li2016a,Marcellina2017}
To demonstrate the $k-$dependence of the Zeeman spin splitting, we tune the front- and backgate electric fields $(\overrightarrow{F}_{\mathrm{front}}, \overrightarrow{F}_{\mathrm{back}})$ on the quantum well to alter $p$ (and hence $k_F$) while keeping the wave function in the quantum well center [Fig. \ref{fig:tilted_field_shubs}(a)]. That is, we set the quantum well to be inversion-symmetric, so that the net electric field $\overrightarrow{F} \equiv F_z \hat{z} \equiv (\overrightarrow{F}_{\mathrm{front}} + \overrightarrow{F}_{\mathrm{back}})/2$ across the quantum well is zero, and there is no Rashba spin-orbit interaction \cite{Rashba1984,Winkler2003}, see Section S1 of the Supplemental Material \footnote{See Supplemental Material at [URL] for details on tuning the quantum well asymmetry, which includes Refs. \cite{Dresselhaus1955,Coleridge1989,Winkler2000,Keppeler2002,Habib2009,Li2016a,Marcellina2017}}. We then introduce an in-plane magnetic field $B_{\parallel}$ to cause an in-plane Zeeman spin splitting. To measure the change in spin splitting as a function of $B_{\parallel}$, we apply a small $B_z$ to cause Shubnikov-de Haas oscillations, from which we measure the area of the Fermi surface. To this end, we tilt the sample at an angle $\theta$ with respect to the total magnetic field $\boldsymbol{B}$ [Fig. \ref{fig:tilted_field_shubs}(b)]. The Shubnikov-de Haas oscillations at various $\theta$ for three different densities are shown in Figs. \ref{fig:tilted_field_shubs}(c)-(e). At $\theta = 0^{\circ}$, for $0 \leq B_z \leq 0.25$ T, there is only one period of magnetoresistance oscillations $\rho_{xx}$, corresponding to a single spin-degenerate Fermi surface, where $\rho_{xx}$ minima occuring only at even filling factors $\nu_{\mathrm{even}}$. At higher $B_z$, the out-of-plane Zeeman splitting $\propto g_{zz} B_z$ becomes visible, with $\rho_{xx}$ minima also developing at odd filling factors $\nu_{\mathrm{odd}}$. Applying an in-plane magnetic field lifts the spin degeneracy even at low $B_z,$ so that $\rho_{xx}$ varies as a function of $\theta$ at a fixed $B_z$ (and hence filling factor $\nu$). The dashed box in Fig. \ref{fig:tilted_field_shubs}(c) shows that the resistance maximum for $p = 1 \times 10^{11}$ cm$^{-2}$ at $\nu = 15$ and $\theta = 0^{\circ}$  evolves into a minimum as $\theta$ is gradually increased. Comparison between Figs. \ref{fig:tilted_field_shubs}(d) and (e) shows that as $p$ is increased, progressively smaller $\theta$ are required to cause this evolution. For $p = 2 \times 10^{11}$ cm$^{-2}$, the resistance minimum at $\nu = 28$ and $0^{\circ}$ evolves into a weak minimum when $\theta$ is increased to $78.6^{\circ}$  [see the dashed box in Fig. \ref{fig:tilted_field_shubs}(d)]. By contrast, for $p = 3 \times 10^{11}$ cm$^{-2}$ [see the dashed box in Fig. \ref{fig:tilted_field_shubs}(e)], a weak minimum already develops at $\theta=63.5^{\circ}$ at $\nu = 28$. This comparison provides evidence that the in-plane Zeeman splitting is density-dependent. In fact, it was predicted that the in-plane Zeeman splitting in heavy holes, up to second order perturbation theory, is given by \cite{Winkler2003}:
\begin{equation}
\begin{array}[b]{rcl}
E(k_{F\pm}) & = &\frac{\hbar^2}{2m^*} k_{F\pm}^2 \pm Z \mu_B  k_{F\pm}^2 B_{\parallel} \\
& \equiv & \frac{\hbar^2}{2m^*} k_{F\pm}^2 \pm \frac{g_{\parallel}(k_{F\pm})}{2} \mu_B B_{\parallel}. 
\end{array}
\label{eq:in_plane_Zeeman_dispersion}
\end{equation}
where $\mu_B$ is the Bohr magneton and the Zeeman prefactor $Z$ is detailed in Section S4.B of the Supplemental Material \footnote{See Supplemental Material at [URL] for additional details and calculations, which includes Ref. \cite{Baldereschi1973}}.  Thus, the effect of $B_{\parallel}$ is equivalent to changing the effective mass for the two spin subbands. 

\begin{figure*}
	\begin{center}
		\includegraphics[scale=0.67,trim={0 1.4cm 0 0}]{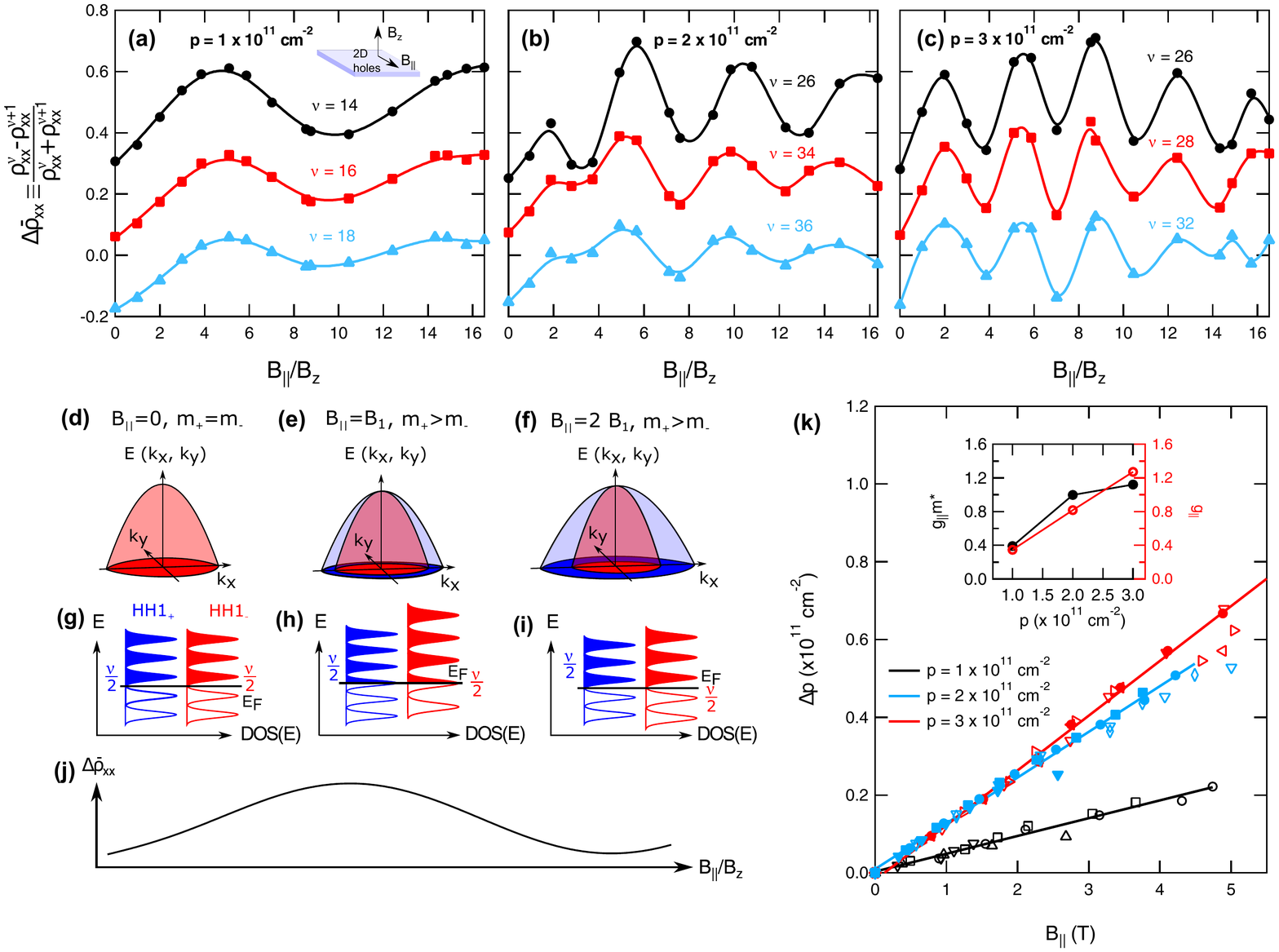}
	\end{center}
	\caption{(a)-(c) Resistance oscillations $\Delta \bar{\rho}_{xx} = \Delta \bar{\rho}_{xx}(B_{\parallel}) \equiv {(\rho_{xx}^{\nu} - \rho_{xx}^{\nu+1})}/{(\rho_{xx}^{\nu} + \rho_{xx}^{\nu+1})}$ at various filling factors $\nu$ for different $p$. (d)-(f) Schematic of the energy band dispersion $E(k_x,k_y)$, showing how the $k$-dependent Zeeman splitting, where $k \equiv \sqrt{k_x^2 + k_y^2}$, changes the effective masses $m_+$ and $m_-$ in the two bands. (g)-(i) Schematic of the Landau levels corresponding to the band structure in (d)-(f) for an even $\nu$. The shaded regions refer to the filled Landau levels.  As $B_{\parallel}$ is increased from $0$ T to $B_{\parallel} = B_{1} > 0$ T, the Landau level spacing in the two spin-split bands changes due to the different masses. (j) The in-plane magnetoresistance $\Delta \bar{\rho}_{xx}$ arises from the change in the density of states at the Fermi energy as illustrated in (d)-(f). (k) The difference in the hole density of the two spin split bands $\Delta p$ as a function of $B_{\parallel}$ for different $p$, as indicated by the three different colours. The inset shows  $g_{\parallel} m^*$ (solid black line with solid circles) extracted using Eq. \ref{eq:in_plane_Zeeman_dispersion}. Using the effective masses obtained from $6 \times 6$ $\boldsymbol{k.p}$ calculations \cite{Birner2007}, we find $g_{\parallel}$ (solid red line with open circles), which increases linearly with density.}
	\label{fig:Rxx_vs_cot_angle}
\end{figure*}

In 2D electrons, the $g-$factor can be extracted from tilted magnetic field measurements by finding the angle $\theta_c$ at which the Zeeman energy $g \mu_B B$ equals half the cyclotron energy  $(1/2) \hbar e B \cos \theta_c /m^{*}$ \cite{Fang1968}. Assuming the electron effective mass $m^*$ is known and $g$ is isotropic ($g_{zz} = g_{\parallel}$), the $g-$factor is given by $g =  (m_0 \cos \theta_c/m^*)$, where $m_0$ is the bare electron mass. However, this approach is not applicable for holes as the $g-$factor is highly anisotropic ($g_{zz} \gg g_{\parallel}$), and $m^*$ depends on $B_{\parallel}$ (Eq. \ref{eq:in_plane_Zeeman_dispersion}, see also Section S2 of the Supplemental Material \footnote{See Supplemental Material at [URL] for detailed calculations.}). Instead, here we examine the dependence of the magnetoresistance on $B_{\parallel}$ at a fixed $B_z$, and hence a fixed $\nu$, to measure the area of the spin-split Fermi surfaces. To increase the visibility of the $B_{\parallel}-$induced magnetoresistance features against the smooth background, we define a dimensionless in-plane magnetoresistance $\Delta \bar{\rho}_{xx}$ at a given $\nu$
\begin{equation}
	\Delta \bar{\rho}_{xx} = \Delta \bar{\rho}_{xx}(B_{\parallel}) \equiv \frac{\rho_{xx}^{\nu} - \rho_{xx}^{\nu + 1}}{\rho_{xx}^{\nu} + \rho_{xx}^{\nu + 1}}.
\end{equation} 
The magnetoresistance  $\bar{\rho}_{xx}$ [Fig. \ref{fig:Rxx_vs_cot_angle}(a)-(c)] oscillates as a function of $B_{\parallel}$, with the frequency of the oscillations being independent of $\nu$. Furthermore, the frequency increases with $p$.

The physical mechanism underlying the density-dependence of $\bar{\rho}_{xx}$ is illustrated in Figs. \ref{fig:Rxx_vs_cot_angle}(d)-(f). We first consider the energy dispersion $E(k_x,k_y)$ in the absence of $B_z$.  When $B_{\parallel} = 0$, the spin subbands HH1$_\pm$ are degenerate [Fig. \ref{fig:Rxx_vs_cot_angle}(d)].  The effect of $B_{\parallel}$ on the energy dispersion is remarkably different for 2D holes than electrons. For 2D electrons, $B_{\parallel}$ splits all $k$ states by $\Delta E_Z = g \mu_B B_{\parallel}$, whereas for 2D holes, $B_{\parallel}$ causes a $k-$dependent splitting  of the HH1$_+$ and HH1$_-$ bands, so that $m_+ \neq m_-$ [Fig. \ref{fig:Rxx_vs_cot_angle}(e)], where $m_{\pm}$ is the effective mass of the HH1$_{\pm}$ bands. The $k$-dependent spin splitting increases with $B_{\parallel}$, and the corresponding effective masses $m_{+}$ and $m_{-}$ diverge further [Fig. \ref{fig:Rxx_vs_cot_angle}(f)] with $B_{\parallel}$. When $B_z$ is finite, Landau levels form, where the energy separation between the Landau levels depends on the effective mass, and hence on $B_{\parallel}$. When $B_{\parallel} = 0$, the HH1$_{\pm}$ Landau levels are spin degenerate, $m_+ = m_-$, and the number of occupied Landau levels is identical in the HH1$_{+}$  and HH1$_{-}$ subbands ($\nu_{+} = \nu_{-}$)  [Fig. \ref{fig:Rxx_vs_cot_angle}(g)]. When $B_{\parallel} \neq 0$, $m_+ $  diverges from $m_-$, changing the spacing of Landau levels in the HH1$_{+}$  and HH1$_{-}$ bands. Thus, the density of states at the Fermi energy $E_F$ as  $\nu_{+}$ and $\nu_-$ changes with $B_{\parallel}$ [Figs. \ref{fig:Rxx_vs_cot_angle}(h)-(i)]. Assuming that the change $\Delta \nu_{\pm}$ in the occupancy of the HH1$_{\pm}$ Landau levels is much smaller than $\nu$ ($\Delta \nu_{\pm}/\nu \ll 1$), the Landau level occupation changes in pairs, i.e. $\Delta \nu_{+} = -\Delta \nu_{-}$ [Fig. \ref{fig:Rxx_vs_cot_angle}(g)].

The change in the density of states at $E_F$ causes $\Delta\bar{\rho}_{xx}$ to oscillate as a function of $B_{\parallel}$ at a fixed $B_z$. When $E_F$ coincides (does not coincide) with a Landau level, a maximum (minimum) in $\Delta\bar{\rho}_{xx}$ develops [Fig. \ref{fig:Rxx_vs_cot_angle}(j)]. Since one oscillation period of $\Delta\bar{\rho}_{xx}$ corresponds to $|\Delta \nu_{\pm}| = 1$, the spin splitting after moving $n$ $\bar{\rho}_{xx}(B_{\parallel})$ resistance peaks away from the symmetry point is given by:
\begin{equation}
	\label{eq:peak_counting}
		\frac{\nu/2 + n \Delta \nu_{\pm}}{\nu} = \frac{p_{\pm}}{p},
\end{equation}
where $p_{\pm}$ is the spin-split densities. Note that although Eq. \ref{eq:peak_counting} is exact for parabolic and isotropic bands (see Section S2 of the Supplemental Material \cite{Note3}), it also holds for non-parabolic and/or anisotropic bands as long as $\Delta \nu_{\pm}/\nu \ll 1$ (see Sections S3 and S4 of the Supplemental Material \footnote{See Supplemental Material at [URL] for a detailed derivation of Eq. \ref{eq:peak_counting}, which includes Ref. \cite{Gunawan2004}}). We remark that the resistance oscillations $\Delta\bar{\rho}_{xx}(B_{\parallel})$ are analogous to Shubnikov-de Haas oscillations: in conventional Shubnikov-de Haas oscillations, the Landau level spacing $e B_z/m^*$ is controlled by varying $B_z$, whereas the oscillations $\Delta\bar{\rho}_{xx}(B_{\parallel})$ are caused by varying $m_{\pm}$. 

We use Eq. \ref{eq:peak_counting} to extract from $\Delta \bar{\rho}_{xx}(B_\parallel)$ the change in the area of the spin-split Fermi surfaces, and hence the spin splitting $\Delta p$,  as a function of $B_\parallel$. We obtain $\Delta p$ for multiple values of $\nu$, as indicated by the symbols in Fig.  \ref{fig:Rxx_vs_cot_angle}(k). The spin splitting is independent of $\nu$, and increases linearly with $B_{\parallel}$. To obtain $g_{\parallel} = g_{\parallel}(k_F)$ from $\Delta p$, we use the dispersion relation in Eq. \ref{eq:in_plane_Zeeman_dispersion} and obtain $Z \mu_B = 1.37,~1.88,~1.95\times10^{-18}$ meV m$^{2}$ T$^{-1}$  and ${g_{\parallel}(k_F) m^*/m_0} = 0.39$, $0.99$, and $1.18$, for $p = 1$, $2$, and $3\times10^{11}$ cm$^{-2}$, respectively (shown by the solid black line and solid circles in the  inset to Fig. \ref{fig:Rxx_vs_cot_angle}(k)). Using the effective masses obtained from $6 \times 6$ $\boldsymbol{k.p}$ calculations \cite{Birner2007}, i.e. $m^* = 1.13, 1.22,$ and $ 0.88 $ $m_0$ for $p = 1,~2,$ and $3 \times 10^{11}$ cm$^{-2}$ respectively, we find $g_{\parallel} = 0.34$, $0.82$, and $1.26$ (shown by the solid red line and open circles in the inset to Fig. \ref{fig:Rxx_vs_cot_angle}(k)). These values of $g_{\parallel}$ are of the same order of magnitude as the predicted values for 2D hole systems \cite{Kernreiter2013,Miserev2017a}. It is also interesting to note that the values of $g_{\parallel}$ we measure here are consistent with the experimental results for quasi-one dimensional hole systems \cite{Nichele2014,Srinivasan2017,Miserev2017b}. 

\begin{figure}
	\centering
	\includegraphics[scale=0.9]{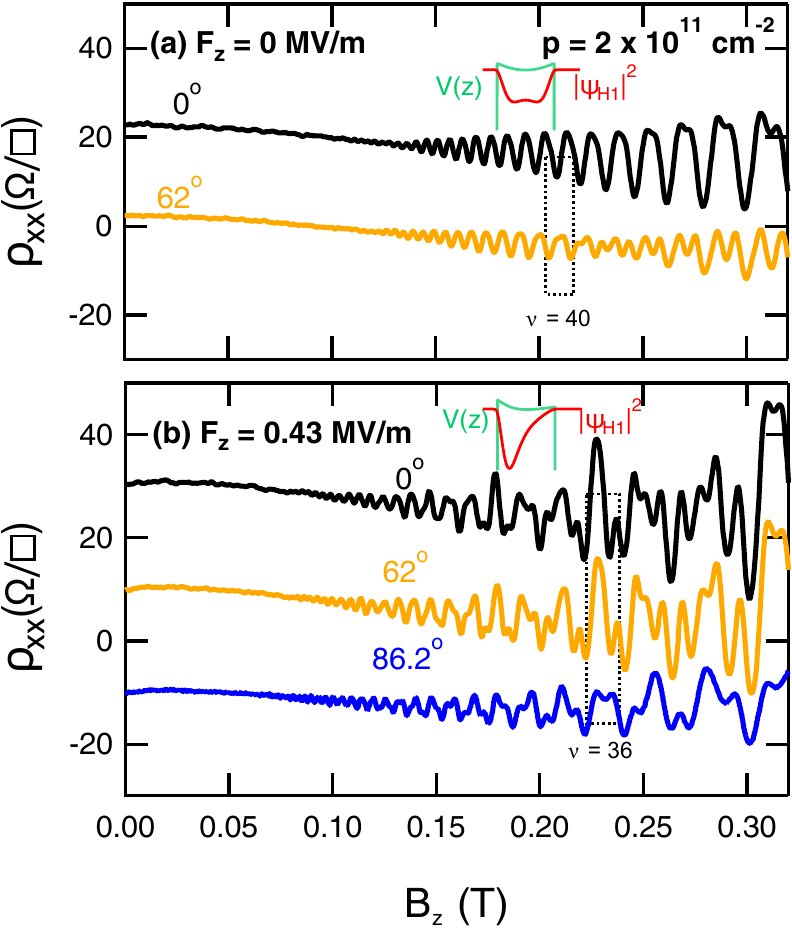}
	\caption{(a) In the absence of Rashba splitting ($F_z = 0$ MV/m), the magnetoresistance minima change into maxima (or vice versa) when $\theta$ is increased from $0^{\circ}$ to $62^{\circ}$, as highlighted in the dashed box for $\nu = 40$. (b) At $F_z = 0.43$ MV/m, the Shubnikov-de Haas oscillations are far less affected by $\theta$: when $\theta$ is increased from $0^{\circ}$ to $62^{\circ}$ the positions of the minima and maxima remain almost unchanged, and only when $\theta$ is large ($\theta > 80^{\circ}$), do the resistance minima evolve into maxima (the dashed box highlights $\nu = 36$). Here, $p = 2 \times 10^{11}$ cm$^{-2}$. The shape of the potential profile and ground state envelope wavefunction are shown in green and red. The traces are offset for clarity.}
	\label{fig:tilted_field_Shubs_non_zero_Fz}
\end{figure}

Finally, we investigate the interplay of Rashba and Zeeman interactions by repeating the tilted magnetic field measurements with a finite electric field $F_z$ applied across the quantum well and a fixed density of $p = 2 \times 10^{11}$ cm$^{-2}$. Figs. \ref{fig:tilted_field_Shubs_non_zero_Fz}(a) and (b) show Shubnikov-de Haas oscillations for $F_z = 0$ MV/m and  $F_z = 0.43$ MV/m. When the quantum well is symmetric ($F_z$ = 0 MV/m), the Rashba splitting is zero, and the magnetoresistance at $\theta=0^{\circ}$ shows a single oscillation period at low $B_z$, with each Landau level being doubly degenerate. In this case, $B_{\parallel}$ has a strong influence on the Shubnikov-de Haas oscillations, as shown by the dashed box. By contrast, when $F_z$ is large ($F_z = 0.43$ MV/m), the magnetoresistance at $\theta=0$ shows a beating due to the Rashba splitting even at $B_\parallel=0$. A Fourier transform of the $\theta = 0^{\circ}$ oscillations shows a large Rashba splitting of $\Delta p / p = 40\%$ (see Section S1 of the Supplemental Material \cite{Note1}). This Rashba interaction dramatically suppresses the effects of $B_{\parallel}$ on the Shubnikov-de Haas oscillations: the positions of maxima and minima for $\theta = 0^{\circ}$ and $\theta = 62^{\circ}$ are practically identical and only at high tilt angles, $\theta > 80^{\circ}$, do the oscillation minima evolve into maxima or vice versa. 
	
It is challenging to perform a complete quantitative analysis to extract $\Delta p$ and $g_{\parallel}$ when both Rashba and Zeeman interactions are finite. The method of counting the periodic Landau level population/depopulation becomes invalid when $\Delta \nu / \nu \ll 1$ is violated, which is the case with $F_z = 0.43$ MV/m. There is currently no analytical or numerical modeling of 2D holes under the Rashba spin-orbit interaction in tilted magnetic field configuration against which we can compare our data. Such calculations are extremely challenging due to the combination of quantum confinement, the four-component hole spinor \cite{Winkler2005}, heavy hole-light hole coupling, as well as the simultaneous inclusion of $B_z$ and $B_{\parallel}$ \cite{Yeoh2014}. To get a qualitative understanding of Fig. \ref{fig:tilted_field_Shubs_non_zero_Fz}, we present a minimal model for the spin splitting when both Rashba and in-plane Zeeman interactions are present at $B_z = 0$ T (see Section S5 of the Supplemental Material \cite{Note3}). The suppression of $g_{\parallel}$ in Fig. \ref{fig:tilted_field_Shubs_non_zero_Fz} is due to the Rashba interaction dominating over the in-plane Zeeman interaction. In this limit,  the effect of $B_{\parallel}$ is to offset the center of the Fermi surfaces from $k = 0$, without changing the areas appreciably. This is because the Rashba and in-plane Zeeman interactions have similar functional forms, i.e. $(k_{-}^3 \sigma_+ + k_{+}^3 \sigma_-)$ and $(k_{-}^2 B_{-} \sigma_+ + k_{+}^2 B_{+} \sigma_-)$, respectively. We note that in the opposite limit, where the in-plane Zeeman interaction is much larger than the Rashba interaction, $g_{\parallel}$ reverts to its value at zero Rashba interaction (Fig. S3(a) of the Supplemental Material \cite{Note3}).

In summary, we have demonstrated all-electrical control of the in-plane Zeeman splitting of 2D holes by separately varying the 2D hole density $p$ and the electric field $F_z$ across the quantum well. We have developed a novel method to quantify the Zeeman splitting from the tilted magnetic field measurements. This new method can in principle be generalized to other 2D materials with an anisotropic $g-$factor ($g_{\parallel} \neq g_{zz}$) as long as the confinement potential is inversion-symmetric. By tracking the evolution of the Landau levels as a function of $B_{\parallel}$, we show that the in-plane Zeeman splitting of 2D holes is proportional to $p$, and that $g_{\parallel}$ can be tripled. However, the experimental and calculated values of $g_{\parallel}$ diverge by up to a factor of 3, which is likely due to exchange and correlations. We have also shown that a strong $F_z$ can suppress $g_{\parallel}$, although a complete quantitative analysis for extracting $g_{\parallel}$ in the case of finite $F_z$ is beyond the scope of the present work. The ability to electrically tune $g_{\parallel}$ will be useful for designing spin-based devices, such as spin-transistors, spin-orbit qubits, quantum logic gates, and hybrid superconductor-semiconductor systems hosting Majorana modes.

\section*{Acknowledgements}

The authors thank Jo-Tzu Hung and Weizhe Liu for enlightening discussions. This work was supported by the Australian Research Council under the DP scheme. We also acknowledge financial support from the EPSRC, UK, author I.F. from Toshiba Research Europe, and author A.F.C. from Trinity College, Cambridge.

%\bibliography{refs}


\end{document}

\title{Supplemental material for ``Electrical control of the Zeeman spin splitting in two-dimensional hole systems"}
	
	\author{E. Marcellina}
	\affiliation{School of Physics, The University of New South Wales, Sydney, Australia}
			
	\author{A. Srinivasan}
	\affiliation{School of Physics, The University of New South Wales, Sydney, Australia}
	
	\author{D. S. Miserev}
	\affiliation{School of Physics, The University of New South Wales, Sydney, Australia}
		
	\author{A. F. Croxall}
	\affiliation{Cavendish Laboratory, University of Cambridge, J. J. Thomson Avenue, Cambridge CB3 0HE, United Kingdom}
	
	\author{D. A. Ritchie}
	\affiliation{Cavendish Laboratory, University of Cambridge, J. J. Thomson Avenue, Cambridge CB3 0HE, United Kingdom}
	
	\author{I. Farrer}
	\affiliation{Cavendish Laboratory, University of Cambridge, J. J. Thomson Avenue, Cambridge CB3 0HE, United Kingdom}
	
	\author{O. P. Sushkov}
	\affiliation{School of Physics, The University of New South Wales, Sydney, Australia}
		
	\author{Dimitrie Culcer}
	\affiliation{School of Physics, The University of New South Wales, Sydney, Australia}
	
	\author{A. R. Hamilton}
	\affiliation{School of Physics, The University of New South Wales, Sydney, Australia}

\maketitle

\section{Tuning the quantum well asymmetry}
\label{sec:QW_asymmetry}

\begin{figure*}
	\begin{center}
		\includegraphics[scale=0.97]{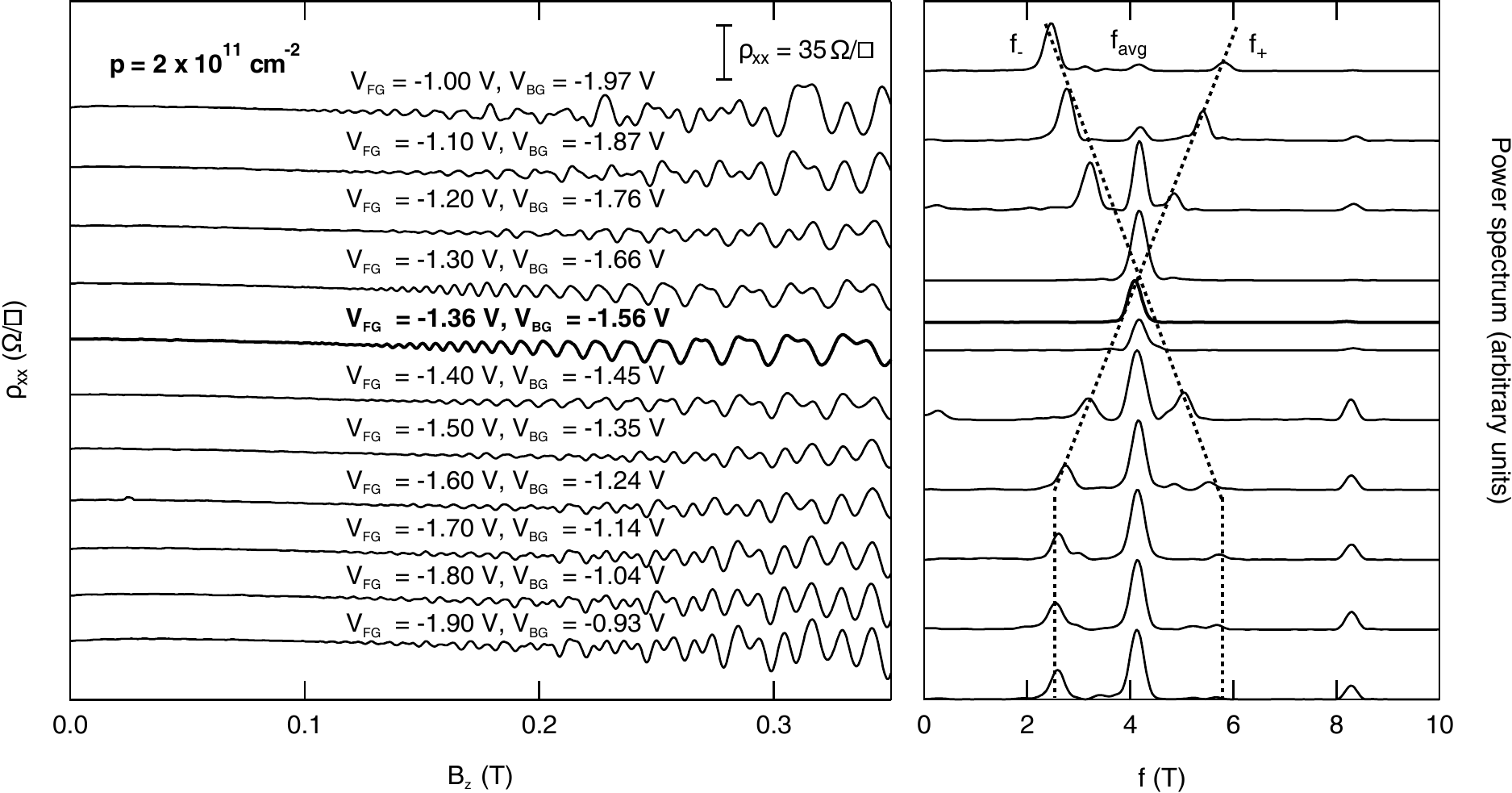}
		\caption{Shubnikov-de-Haas oscillations for different gate settings ($V_{\text{FG}}$, $V_{\text{BG}}$), where $V_{\text{FG(BG)}}$ is the front (back) gate voltage, at density $p = 2\times10^{11}$ cm$^{-2}$ and the corresponding Fast Fourier Transform spectra. At the symmetry point (i.e. the oscillations corresponding to $F_z = 0$ MV/m, highlighted by the bold traces), the Shubnikov-de Haas oscillations show one period and only one Fourier peak at 4.18 T is resolved. The measurements in Figs. 1d and 3a of the main text are performed at the symmetry point. Away from the symmetry point, spin splitting occurs, as shown by the dashed lines. The Fourier peaks at $f_{\pm}$ correspond to the spin-split subbands $p_{\pm}$, whilst the $f_{\mathrm{avg}} = (f_+ + f_-)/2$ peak at 4.18 T is due to non-adiabatic spin dynamics \cite{Winkler2000,Winkler2003,Keppeler2002,Habib2009,Li2016,Li2016a}.  The topmost Shubnikov-de Haas oscillations and Fourier spectra correspond to vertical electric field $F_z = 0.43$ MV/m in Fig. 3b of the main text. The position of the  $f_{\mathrm{avg}}$ peak remains unchanged at 4.18 T, indicating that the hole density was constant throughout. }
		\label{fig:Vary_Rashba_and_FFT_2e15}
	\end{center}
\end{figure*} 

In this section, we explain in detail how we tuned the quantum well to be inversion-symmetric (Fig. 1 of the main text) and varied the quantum well asymmetry at a given density (Fig. 3 of the main text). To this end, we measured the Shubnikov-de Haas oscillations at a fixed density and calculated the spin splitting. From the spin splitting data, we were able to associate the front- and backgate voltages to the electric field $F_z$ across the quantum well.
	
Fig. \ref{fig:Vary_Rashba_and_FFT_2e15} shows the Shubnikov de-Haas oscillations for density $p = 2 \times 10^{11}$ cm$^{-2}$. The beating patterns in the Shubnikov-de Haas oscillations were caused by the Rashba spin-orbit interaction that arises when the quantum well was inversion-asymmetric \cite{Rashba1984,Winkler2000,Winkler2003,Habib2009,Nichele2014}. To resolve the spin splitting, a Fourier analysis was performed on the Shubnikov-de Haas oscillations. To this end, we need to first single out the magnetoresistance oscillations from the smooth background and the decay envelope.  The low-field magnetoresistance oscillations $\Delta \rho_{xx}$  of an $s-$subband system are described as \cite{Coleridge1989}
\begin{widetext}
	\begin{equation}
	\Delta \rho_{xx} \propto \sum_s \frac{2\pi^2 k_B T/\hbar \omega_{s}}{\sinh(2\pi^2 k_B T/\hbar \omega_{s})} \exp \left({-\frac{\pi}{\omega_{s} \tau_{s}}}\right)  \cos \left(\frac{2 \pi E_F}{\hbar \omega_{s}} - \pi \right), 
	\end{equation}
\end{widetext}
where $k_B$ is the Boltzmann constant, $T$ is the temperature, $\tau_{s}$ is the quantum lifetime of the $s-$th subband, $\hbar\omega_{s}$ is the cyclotron energy of the $s-$th subband, and the period of the magnetoresistance oscillations due to the $s-$th subband is $\frac{\hbar \omega_{s}}{2 \pi E_F}$. To isolate the term $\cos \left(\frac{2 \pi E_F}{\hbar \omega_{s}} - \pi \right)$ from the Shubnikov-de Haas oscillations in Fig. \ref{fig:Vary_Rashba_and_FFT_2e15}, three steps were taken. In the first step, the traces in \ref{fig:Vary_Rashba_and_FFT_2e15} were interpolated with respect to $1/B_{z}$ so that the oscillations $\rho_{xx}$ are periodic in $1/B_{z}$. In the second step, the traces $\rho_{xx}(1/B_{z})$ were subtracted by a smooth background, i.e. a sixth-degree polynomial fit to center the results $\Delta \rho_{xx}(1/B_{z})$ around $\Delta \rho_{xx}(1/B_{z}) = 0$. In the third step, the traces $\Delta \rho_{xx}(1/B_{z})$ were normalized by $\exp(0.3/B_{z})$ to cancel the envelope $\exp \left({-\frac{\pi}{\omega_{s} \tau_{s}}}\right) $ of the Shubnikov-de Haas	oscillations, so that the periodic oscillations are directly proportional to $\cos \left(\frac{2 \pi E_F}{\hbar \omega_{s}} - \pi \right)$. The resulting traces $\Delta \rho_{xx} \exp{(0.3/B_{z})}$ were then Fourier transformed using the Fast Fourier Transform routine in Igor Pro and the corresponding Fourier spectra are shown in the insets to Figure \ref{fig:Vary_Rashba_and_FFT_2e15}. The range for the Fast Fourier Transform was chosen to be from $B_{z} = 0.135$ T to $B_{z} = 0.262$ T, since the Rashba spin splitting is more clearly seen at low magnetic fields $B_{z}$ where the out-of-plane Zeeman splitting $g_{zz}B_{z}$ is negligible. 

The Fourier spectra corresponding to the inversion-asymmetric configuration are remarkably different from the symmetric one. When the quantum well is inversion-asymmetric ($F_z \neq 0$ MV/m), three Fourier peaks are resolvable, where the $f_{\pm}$ peaks correspond to the spin-split subbands. The peak at $f_{\mathrm{avg}} = 4.18$ T when the quantum well is asymmetric arises from non-adiabatic spin dynamics that prevails when the strength of the spin-orbit interaction is insufficient to align the spins with the spin-orbit induced effective magnetic field \cite{Winkler2000,Winkler2003,Keppeler2002,Habib2009,Li2016,Li2016a}. We note that the topmost trace in Fig. \ref{fig:Vary_Rashba_and_FFT_2e15} corresponds to a vertical electric field $F_z = 0.43$ MV/m in Fig. 3b of the main text. When the quantum well is inversion-symmetric instead ($F_z = 0$ MV/m), there is only one period of oscillations and hence only one Fourier peak at $f_{\mathrm{avg}} = 4.18$ T develops. The measurements in Figs.  1d and 3a of the main text were performed at $F_z = 0$ MV/m, i.e. at the gate settings where the corresponding Shubnikov-de Haas oscillations show the least beating (bold traces in Fig. \ref{fig:Vary_Rashba_and_FFT_2e15}). Finally, we note that the position of the $f_{\mathrm{avg}} = 4.18$ T remains unchanged with the gate settings, which indicates that the total hole density was constant throughout. 

We repeated the above measurements for other densities $p = 1 \times 10^{11}$ cm$^{-2}$ and $p = 3 \times 10^{11}$ cm$^{-2}$, and find the quantum well can also be tuned symmetrically at these densities. We note that for 2D holes in zincblende materials the Dresselhaus spin-orbit interactions predominantly go as $k^3$ \cite{Dresselhaus1955}, however, we find that such interactions were negligible in our sample even at high densities ($p = 3 \times 10^{11}$ cm$^{-2}$), consistent with recent calculations for 2D holes \cite{Marcellina2017}. This was indicated by the fact that for the densities considered in this work, there were always gate settings corresponding to which only one Fourier peak was resolvable.

The tilted magnetic field measurements were carried out with the quantum well set in the inversion-symmetric configuration at various densities $p = 1,~2,~3 \times 10^{11}$  cm$^{-2}$ [Figs. 1(c)-(e) of the main text] or in the inversion-asymmetric configuration at a fixed density $p = 2 \times 10^{11}$ cm$^{-2}$ (Fig. 3 of the main-text). Using a self-consistent Poisson-Schr\"odinger solver \cite{Birner2007}, the confinement potential $V(z)$ can be calculated and the electric field $F_z$ can be obtained using the relation $F_z = \left[-\frac{dV(z)}{dz}|{_\text{front}} - \frac{dV(z)}{dz}|{_\text{back}}\right]/2$, where $-\frac{dV(z)}{dz}|{_\text{front(back)}}$ is the electric field induced by the front (back) gate.

\section{Proof of Eq. 3 of the main text for parabolic and isotropic bands}
\label{sec:Zeeman_parabolic_bands}

In this section, we derive the relation between the number of the occupied Landau levels and the Zeeman spin-split densities, with the assumption that the HH$1\pm$ bands are both parabolic and isotropic. Since only two subbands HH1$\pm$ are occupied, the total density is given by $p = p_+ + p_-$, and their difference $\Delta p$ is given by $\Delta p = p_+ - p_-$, where $p_{\pm}$ represent the population in the HH1$\pm$ subband. The change in the HH1$\pm$ population, as an in-plane magnetic field is applied, is given as
\begin{equation}
	\begin{array}[b]{rcl}
  		\frac{\Delta p}{2} &=&  \left(p_{\pm} - p/2 \right) \\
		&=&  \frac{1}{4 \pi} \left(k_{F\pm}^2 - k_{F0}^2 \right) \\
	  	&=&  \frac{1}{2 \pi \hbar^2} \left(E_F m_{\pm} - E_{F0} m^* \right)
	 	\label{eq:population_change}
	\end{array}
\end{equation}
where $E_{F0}$, $k_{F0}$, $m^*$ are the Fermi energy, Fermi wave vector, and effective mass at the symmetry point, respectively. Note that the second line of Eq. \ref{eq:population_change} was obtained from the first line using the relation $p_{\pm}=\frac{k_{F\pm}}{4\pi^2}$, which applies for isotropic bands. Furthermore, since the bands are parabolic, the Landau levels are equally spaced, so that the number $\nu_{\pm}$ of filled HH1$\pm$ Landau levels is given as
\begin{equation}
	\nu_{\pm} = \frac{E_F}{\hbar \omega_{c\pm}} = \frac{E_F m_{\pm}}{\hbar e B_{z}},
\end{equation}
the change $\Delta \nu_{\pm}$ in the number of filled Landau levels from the symmetry point is therefore
\begin{equation}
	\begin{array}[b]{rcl}
	{\Delta \nu_{\pm}} &=& \pm \left(\nu_{\pm} - \nu/2 \right) \\
		&=& \pm \frac{1}{\hbar e B_{z}} \left(E_F m_{\pm} - E_{F0} m_0 \right) \\
		&=& \pm \frac{h}{e B_{z}} \left(p_{\pm} - p/2 \right).
	\label{eq:filling_factors_change}
	\end{array}
\end{equation}
By using the relation $\nu = \nu_{+} + \nu_{-}$, and combining Eq. \ref{eq:population_change} with Eq. \ref{eq:filling_factors_change}, we obtain Eq. 3 in the main text. Note that the third line of Eq. \ref{eq:filling_factors_change} was obtained from the second line using $\frac{p_+}{p_-}=\frac{m_+}{m_-}$, which is valid for isotropic parabolic bands. The latter relation can be straightforwardly verified as follows. From Eq. 1 of the main text, one obtains $p_{\pm} = \frac{p}{2}\pm\frac{Z \mu_B B_{\parallel}}{\hbar^2}p$. Then, by using the standard expression for the density of states effective mass for isotropic bands $\frac{m_{\pm}}{m_0} =  \frac{\hbar^2}{m_0} \left(\frac{1}{k} \frac{\partial E(k_{F\pm})}{\partial k} \right)_{k=k_F}^{-1}$, to the lowest order one obtains $	\frac{m_{\pm}}{m_0} = \frac{m^*}{m_0} \left(1 \pm \frac{2 Z \mu_B B_{\parallel}}{\hbar^2} \right)$, thus proving the relation $\frac{p_+}{p_-}=\frac{m_+}{m_-}$.

Eq. 3 is also exact for systems with a $k-$independent Zeeman splitting $\Delta E_z = g \mu_B B$. This can be verified by replacing $E_F$ in Eqs. \ref{eq:population_change}-\ref{eq:filling_factors_change} with $E_F \pm g \mu_B B$ and $m_{\pm}$ by $m^*$. 

\section{Comments on Eq. 3 of the main text for nonparabolic and/or anisotropic bands}
\label{sec:Zeeman_nonparabolic_bands}

We now consider the case of isotropic nonparabolic bands (such as heavy holes with wave vector close to the anticrossing with higher energy subbands) with dispersion relation $E(k_{F\pm}) = \frac{\hbar^2 k_{F\pm}^2}{2 m^*} \pm f(k_{F\pm})$. The spin-split effective masses are given as: 
\begin{equation}
	\begin{array}[b]{rcl}
	\frac{m_{\pm}}{m_0} &=&  \frac{\hbar^2}{m_0} \left(\frac{1}{k} \frac{\partial E(k_{F\pm})}{\partial k} \right)_{k=k_F}^{-1} \\
	&\approx&  \frac{m^*}{m_0} \left(1 \pm \frac{1}{k_{F\pm}}  \frac{\partial f(k)}{\partial k} \vert_{k_{F\pm}} + \mathcal{O}(k_{F\pm}^2) \right).  
	\label{eq:eff_mass}
	\end{array}
\end{equation}
In the case of small perturbation $|f(k_{F})| \ll \frac{\hbar^2 k_{F\pm}}{2 m^*}$ (or equivalently, $\Delta \nu/\nu \ll 1$), the higher-order terms can be neglected and the second term of Eq. \ref{eq:eff_mass} becomes independent of $k_{F\pm}$, hence, we recover Eq. 3.

For anisotropic parabolic bands, Eq. 3 can be used to evaluate the Zeeman splitting as long as the cyclotron effective mass is defined. For example, for ellipsoidal Fermi contours, the effective mass $m_{\pm}$ in Eqs. \ref{eq:population_change}-\ref{eq:filling_factors_change} can be substituted by $\sqrt{m_l m_t}$, where $m_{l(t)}$ is the effective mass in the longitudinal (transverse) direction \cite{Gunawan2004}, and $E_F$ by $E_F \pm g \mu_B B$.

Eq. 3 can also be used for systems with anisotropic and non-parabolic bands, as long as the effective cyclotron mass is known, and that the condition $\Delta \nu/\nu \ll 1$ is fulfilled. This is because our method of counting the periodic Landau level population/depopulation only measures the change in the area of the Fermi surface \textit{irrespective} of its shape, as long as the change is sufficiently small (see also Subsec. \ref{subsec:cubic_terms}). For example, one may include the newly discovered higher-order Zeeman terms $Z'(k_-^4 B_+\sigma_+ + k_+^4 B_-\sigma_-)$ \cite{Miserev2017a} on top of the quadratic$-k$ Zeeman terms $Z(k_-^4 B_-\sigma_+ + k_+^4 B_+\sigma_-)$ (which gives the dispersion in Eq. 1), and then use Eq. 3 to obtain the Zeeman splitting.

\section{Importance of the cubic term in the in-plane Zeeman interactions}
\label{sec:cubic_terms}

Here we discuss in detail the in-plane Zeeman splitting with the cubic terms included and the in-plane Zeeman splitting within spherical approximation.

\subsection{In-plane Zeeman interactions with the cubic terms}

For two-dimensional holes, to the second order in perturbation expansion the in-plane Zeeman interaction containing the terms with cubic symmetry is given as \cite{Winkler2003}
\begin{equation}
	\begin{array}[b]{rcl}
	H_Z &=& z_{51}(B_x k_x^2 \sigma_x - B_y k_y^2 \sigma_y) + z_{52}(B_x k_y^2 \sigma_x - B_y k_x^2 \sigma_y) \\
	&~& + z_{53}\{k_x,k_y\}(B_y \sigma_x - B_x \sigma_y),
	\end{array}
	\label{eq:Zeeman_with_cubic_terms}
\end{equation}
where $\sigma_{x,y,z}$ are the Pauli matrices, and
\begin{subequations}
	\begin{align}
	z_{51} &= -\frac{3}{2} \kappa \gamma_2 Z_1 + 6 \gamma_3^2 Z_2 \\
	z_{52} &= \frac{3}{2} \kappa \gamma_2 Z_1 - 6 \gamma_2 \gamma_3 Z_2  \\
	z_{53} &= 3 \kappa \gamma_3 Z_1 - 6 \gamma_3(\gamma_2+\gamma_3) Z_2. 
	\end{align}
\end{subequations}
For a symmetric quantum well of width $d$, we have
\begin{subequations}
	\begin{align}
	Z_1 &= \frac{d^2}{\pi^2 \gamma_2} \\[1ex] 
	Z_2 &= \frac{512d^2}{27\pi^4(3\gamma_1+10\gamma_2)},
	\end{align}
	\label{eq:Zeeman_Z_terms}
\end{subequations}
where $\gamma_1,\gamma_2,\gamma_3$ are the Luttinger parameters and $\kappa$ is the bulk $g-$factor \cite{Luttinger1956}. 

\subsection{In-plane Zeeman interactions in the spherical approximation}

In the spherical approximation, the Luttinger parameters $\gamma_2$ and $\gamma_3$ in Eqs. \ref{eq:Zeeman_with_cubic_terms}-\ref{eq:Zeeman_Z_terms} are replaced by $\bar{\gamma}=\frac{2\gamma_2+3\gamma_3}{5}$ \cite{Baldereschi1973}. We then obtain Eq. 1, with 
\begin{equation}
	Z = -\frac{3}{2} \kappa \bar{\gamma} Z_1 + 6 \bar{\gamma}^2 Z_{2,\mathrm{spher.}},
	\label{eq:Z_spher}
\end{equation}
where 
\begin{equation}
	Z_{2,\mathrm{spher.}} = \frac{512d^2}{27\pi^4(3\gamma_1+10\bar{\gamma})}.
	\label{eq:Z2_spher}
\end{equation}

\subsection{Comparing the in-plane Zeeman interactions with the cubic terms and in the spherical approximation}
\label{subsec:cubic_terms}

\begin{figure*}
	\begin{center}
		\includegraphics[scale=0.67]{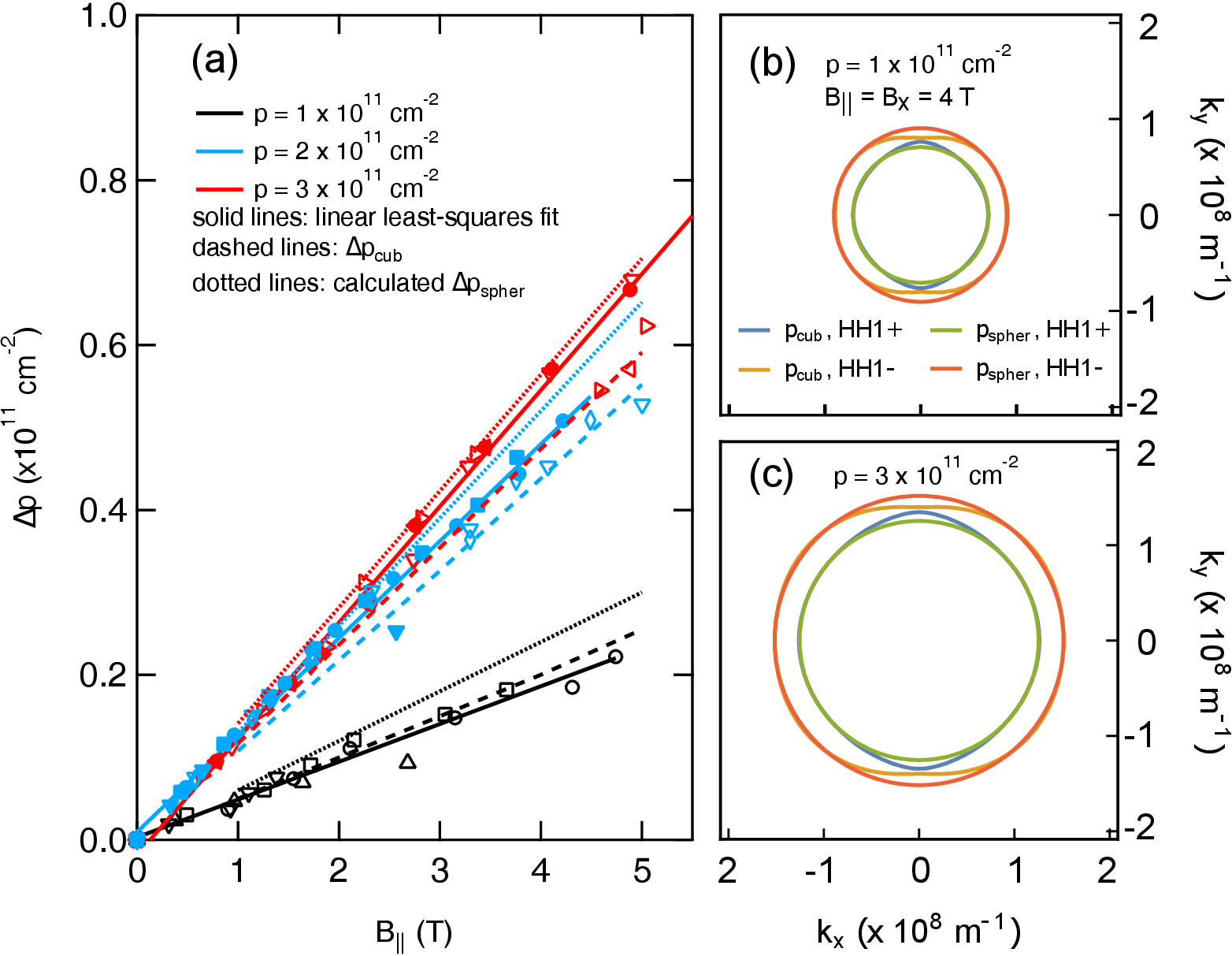}
		\caption{\label{fig:Zeeman} (a) The data in Fig. 2k compared to the in-plane Zeeman splitting $\Delta p_{\mathrm{cub.}}$ calculated with the cubic terms and that within the spherical approximation $\Delta p_{\mathrm{spher}}$. We find that $\Delta_{\mathrm{cub}}$ and $\Delta_{\mathrm{spher}}$ differ by $\lesssim4.5\%$. The Fermi contours for (b) $p = 1 \times 10^{11}$ cm$^{-2}$ and (c) $p = 3 \times 10^{11}$ cm$^{-2}$ at $B_{\parallel} = 4$ T, showing that the cubic terms only affect the Fermi contour in the direction perpendicular to $B_{\parallel}$.}
	\end{center}
\end{figure*}

Here we calculate the in-plane Zeeman splitting $\Delta p$, with both cubic and spherical terms, as a function of $B_{\parallel}$ and compare the results with the experimental data in Fig. 2k [Fig. \ref{fig:Zeeman}(a)]. We find that the in-plane Zeeman splitting $\Delta p_{\mathrm{cub}}$, calculated with the cubic terms taken into account (Eqs. \ref{eq:Zeeman_with_cubic_terms}-\ref{eq:Zeeman_Z_terms}), differs by $\lesssim4.5\%$ from that obtained in the spherical approximation $\Delta p_{\mathrm{spher}}$ (Eqs. 1, \ref{eq:Z_spher}, and \ref{eq:Z2_spher}), i.e. $\frac{|\Delta p_{\mathrm{cub}}-\Delta p_{\mathrm{spher}}|}{(\Delta p_{\mathrm{cub}}+\Delta p_{\mathrm{spher}})/2} \lesssim 4.5\%$, regardless of $p$. We note that most of our experimental data points fall between $\Delta p_{\mathrm{cub}}$ and $\Delta p_{\mathrm{spher}}$, suggesting that the actual Fermi spin-split contours are not perfectly circular, although our initial assumption of isotropic Fermi contours (Eq. 1) proves to be an excellent approximation. Furthermore, the small difference between the cubic and spherical models, as well as their closeness to our experimental data, also indicates that our method of counting the periodic Landau level population/depopulation only probes the change in the area of the spin-split Fermi contours, regardless of their exact shape, as long as the perturbation remains small.

Figs. \ref{fig:Zeeman}(b) and (c) show the Fermi contour of the spin-split bands for $p = 1 \times 10^{11}$ cm$^{-2}$ and $p = 3 \times 10^{11}$ cm$^{-2}$, respectively. We note that the cubic terms distort the Fermi contour in the direction perpendicular to $B_{\parallel}$, while the direction parallel to $B_{\parallel}$ is almost unaffected. We also notice that the cubic corrections do not depend on density, which is likely due to the fact that both the spherical and cubic terms in the in-plane Zeeman interactions go as $\sim k^2$. 

\begin{figure*}
	\includegraphics[scale=1.2,trim={0cm 0.25cm 0cm 0cm}]{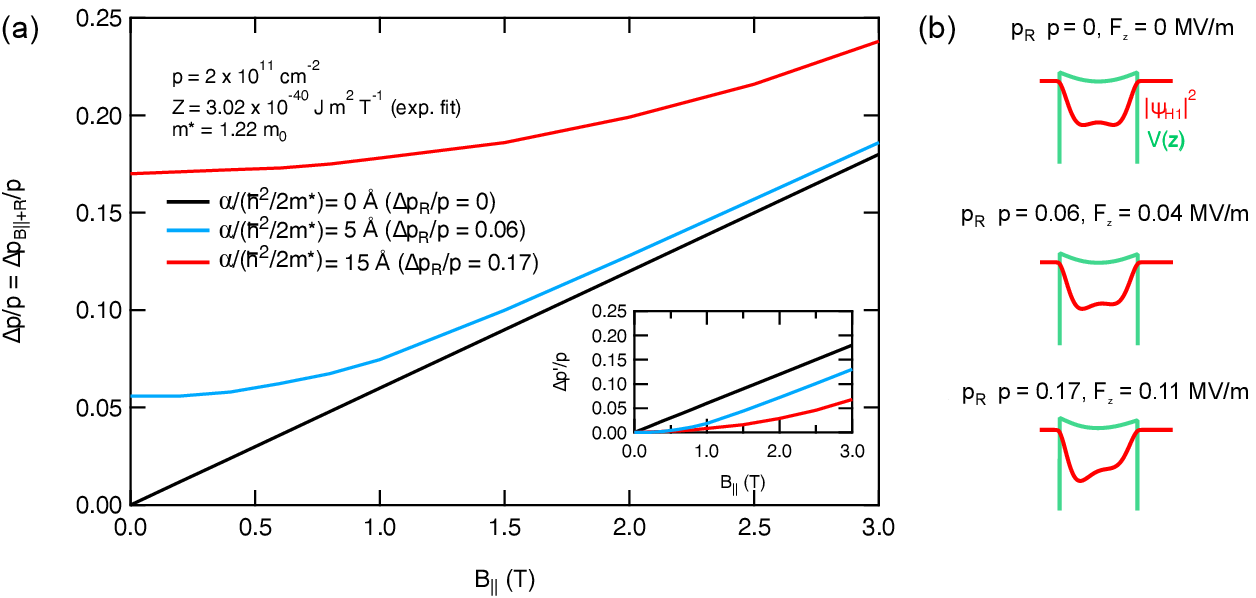}
	\caption{(a) The spin splitting $\Delta p/p = \Delta p_{B\parallel+\mathrm{R}}/p$, calculated using Eq. \ref{eq:spin_splitting} in the case of finite in-plane Zeeman and Rashba interactions. The symbol $\Delta p_R/p$ denotes the Rashba spin splitting at $B_{\parallel} = 0$ T, i.e. $\Delta p_R/p \equiv \Delta p (B_{\parallel} = 0~ \mathrm{T})/p$. Inset: the total spin splitting subtracted by the spin splitting at $B_{\parallel} = 0$ T, i.e. $\Delta p'/p \equiv \Delta p/p - \Delta p_R/p$ as a function of $B_{\parallel}$, showing that at low $B_{\parallel}$ ($R \ll 1$), the Rashba interaction suppresses the in-plane Zeeman splitting. At high $B_{\parallel}$, the Zeeman interaction dominates and the spin splitting is directly proportional to $Z$. (b) Schematic of the confinement potential $V(z)$ for the various strengths of Rashba interaction (again characterized by $\Delta p_R/p$), that arises from the application of a vertical electric field $F_z$.}
	\label{fig:Zeeman_Rashba} 
\end{figure*}

\begin{figure}
	\includegraphics[scale=0.65,trim={0 0cm 0 0}]{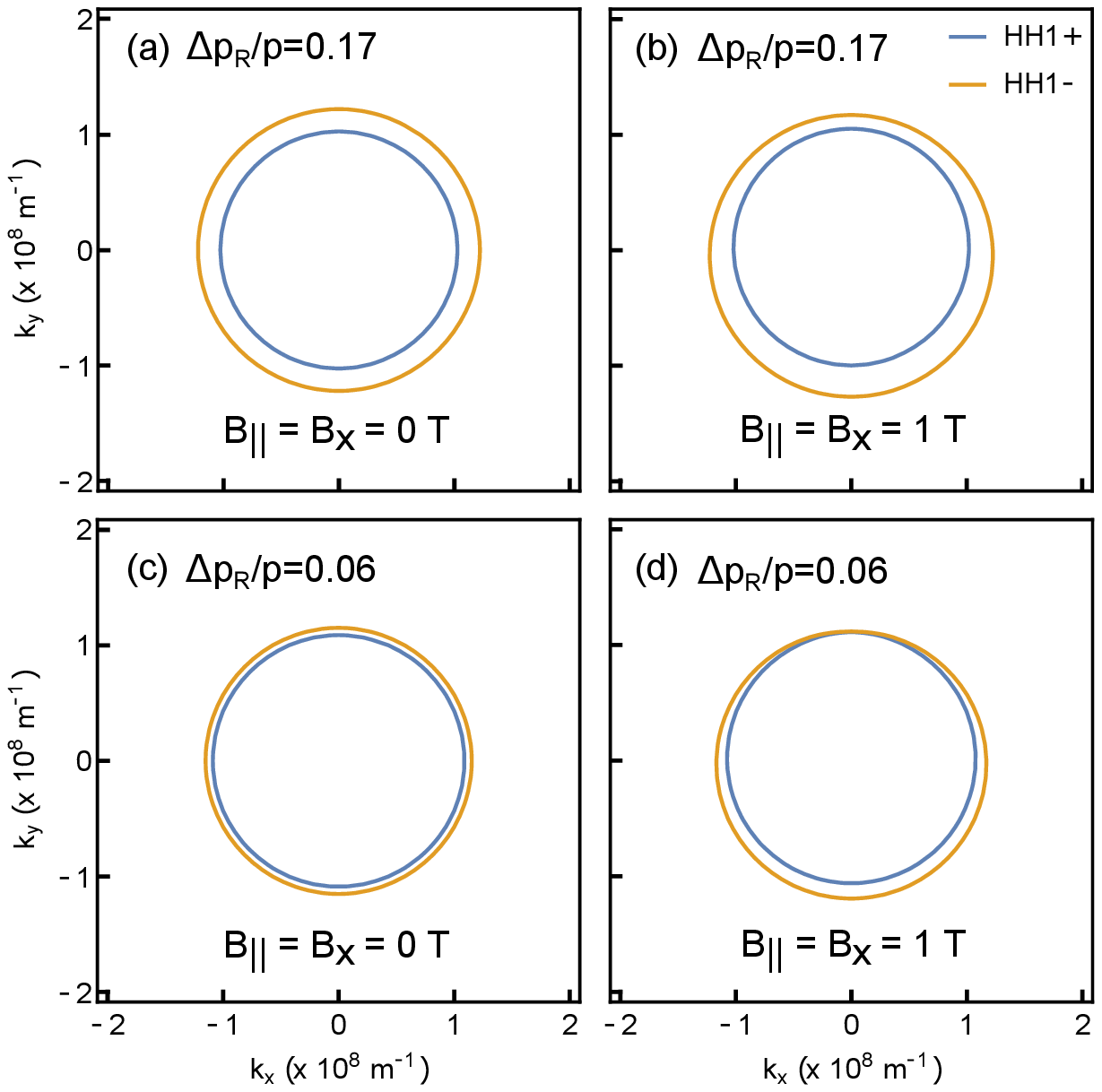}
	\caption{Spin-split Fermi contours when both Rashba and $B_{\parallel}$ is present.  For $\Delta p_R/p=0.17$ at (a) $B_{\parallel} = 0~$T and (b) $B_{\parallel} = 1~$T, and for $\Delta p_R/p=0.06$ at (c) $B_{\parallel} = 0~$T and (d) $B_{\parallel} = 1~$T. Here $p = 2 \times 10^{11}$ cm$^{-2}$ and $m^* = 1.22~ m_0$. When Rashba interaction dominates, the Zeeman interaction moves the spin-split Fermi contours with respect to each other.} 
	\label{fig:Fermi_contours}
\end{figure}

\section{Spin splitting when both Rashba spin-orbit and in-plane Zeeman interactions prevail}
\label{sec:spin_splitting_Rashba_Zeeman}

The tilted magnetic field measurements in Fig. 3 suggests that the Rashba interaction can suppress the in-plane Zeeman splitting. However, the interplay between Rashba and in-plane Zeeman interactions, especially at finite $B_z$, is highly non-trivial due to a combination of factors. Firstly, the Rashba interaction changes the heavy hole-light hole wave function overlaps and the heavy hole-light hole splitting that determine the in-plane $g-$factor. Secondly, the contribution of the Rashba spin-orbit interaction to the spin-splitting itself is highly dependent on density and effective mass. Thirdly, the Rashba and in-plane Zeeman interactions align the spins in different orientations, so that the spin orientation may have to be described by a four-component spinor, as discussed in  Ref. \cite{Winkler2005} for a symmetric quantum well. Finally, modeling the hole band structure with both $B_{\parallel}$ and $B_z$ included is highly impractical due to the aforementioned factors and the large amount of basis functions used for performing such calculations \cite{Yeoh2014}. To the best of our knowledge, there are no existing calculations on hole systems with both Rashba and in-plane Zeeman interactions included, and neither for hole systems under the Rashba spin-orbit interaction in a tilted magnetic field configuration. A full treatment of the combined effects of Rashba and in-plane Zeeman interactions, especially at finite $B_z$, is beyond the scope of this paper and will be of interest for further studies. 

In the following we present a simple model to explain how the spin splitting behaves as a function of $B_{\parallel}$ when both Rashba and in-plane Zeeman interactions prevail. Since we are operating in the regime where the out-of-plane Zeeman splitting $g_{zz} \mu_B B_z$ is negligible, we set $B_z = 0$ T throughout. In this model we also ignore the hole spin orientation and the effects of $B_{\parallel}$ on the orbital degree of freedom. We consider the following Hamiltonian:
\begin{equation}
	H = H_0 + H_{\mathrm{SO}},
	\label{eq:generic_Hamiltonian}
\end{equation}
where $H_0 = \frac{\hbar^2 k^2}{2 m^*}$ is the kinetic term and,
\begin{equation} 
\begin{array}[b]{rcl}  
H_{\mathrm{SO}} &=& Z(k_{-}^2 B_{-} \sigma_+ + k_{+}^2 B_{+} \sigma_-) + \alpha(k_{-}^3 \sigma_+ + k_{+}^3 \sigma_-)\\[1ex]
 &=&  k^2 \left[ (Z B_{\parallel} \cos(2 \theta + \phi) - k \alpha \sin (3 \theta)) \sigma_x \right. \\
&+& \left. (Z B_{\parallel} \sin(2 \theta + \phi) + k \alpha \cos (3 \theta)) \sigma_y  \right], 
\end{array}
\end{equation}  
where $\alpha$ is the Rashba coefficient, $k_{\pm} = k_x \pm i k_y$, $B_{\pm} = B_x \pm i B_y$, $k = \sqrt{k_x^2 + k_y^2}$, $B_{\parallel} = \sqrt{B_x^2 + B_y^2}$, $\theta = \text{arctan}(k_y/k_x)$, $\sigma_{\pm} = \frac{1}{2}(\sigma_x \pm \sigma_y)$ and  $\phi = \text{arctan}(B_y/B_x)$. The eigenenergies of Eq. \ref{eq:generic_Hamiltonian} are
\begin{equation}
E(k,\theta) = E_0 \pm k^2 \sqrt{g(k)}.
\label{eq:spin_splitting}
\end{equation}
with 
\begin{equation}
g(k) \equiv \alpha^2  (1 + R^2 - 2 R \sin(2 (\theta - \phi)))
\end{equation}
and $R \equiv \frac{Z B_{\parallel}}{\alpha k}$.

Fig. \ref{fig:Zeeman_Rashba}(a) shows the total spin splitting $\Delta p/p = \Delta p_{B\parallel+\mathrm{R}}/p$ in the case of finite in-plane Zeeman and Rashba interactions for various strengths of Rashba spin-orbit interaction [which parametrized by the Rashba spin splitting $\Delta p_R/p \equiv \Delta p (B_{\parallel} = 0~ \mathrm{T})/p$]. The inset to Fig. \ref{fig:Zeeman_Rashba}a shows the total spin splitting subtracted by the spin splitting at $B_{\parallel} = 0$ T, i.e. $\Delta p'/p \equiv \Delta p/p - \Delta p_R/p$ as a function of $B_{\parallel}$. Assuming that $\phi = 0$ and $Z$ is constant throughout, we find that:
\begin{enumerate}
	\item At low $B_{\parallel}$ ($R \ll 1$), the slope $[d(\Delta p/p)/d B_{\parallel}](\alpha \neq 0)$  [red and blue traces in Fig. \ref{fig:Zeeman_Rashba}(a)] is significantly reduced compared to $[d(\Delta p/p)/d B_{\parallel}](\alpha = 0)$  [black trace in Fig. \ref{fig:Zeeman_Rashba}(a)]. 
	\item At high $B_{\parallel}$ ($R \gg 1$), the slope reverts to $[d(\Delta p/p)/d B_{\parallel}](\alpha = 0)$, i.e. $d(\Delta p/p)/d B_{\parallel}$ becomes directly proportional to $Z$ [compare the blue and black traces in Fig. \ref{fig:Zeeman_Rashba}(a)].
	\item For $\Delta p_R/p = 0.06$, the Zeeman splitting starts to dominate at $B_{\parallel} \gg 0.9$ T and for $\Delta p_R/p = 0.17$ at $B_{\parallel} \gg 2.7$ T.
\end{enumerate}
The confinement potential corresponding to the strengths of the Rashba interaction in Fig. \ref{fig:Zeeman_Rashba}(a) is shown in Fig. \ref{fig:Zeeman_Rashba}(b).

Fig. \ref{fig:Fermi_contours} shows the Fermi contour for $\Delta p_R/p=0.17$ at (a) $B_{\parallel} = 0~$T and (b) $B_{\parallel} = 1~$T, and for $\Delta p_R/p=0.06$ at (c) $B_{\parallel} = 0~$T and (d) $B_{\parallel} = 1~$T. We find that when Rashba interaction is dominant, the in-plane Zeeman interaction does not contribute much to the spin-splitting, rather, the in-plane Zeeman interaction moves the spin-split Fermi contours with respect to each other.

Based on our simple model, we believe that the suppression of the in-plane $g-$factor $g_{\parallel}$ in Fig.  3(b) of the main text is due to the Rashba interaction overwhelming the in-plane Zeeman interaction. In this regime, the in-plane Zeeman interaction shifts the spin-split Fermi contours with respect to each other, without altering the area of the Fermi contours much. However, a full treatment of Rashba and Zeeman splitting likely requires four-component spinors, and is beyond the scope of this work \cite{Winkler2005}.
		
%\bibliography{refs}
